\documentclass{JAC2003}
\usepackage{graphicx}
\usepackage{booktabs}

\begin{document}
\title{TILTED SEXTUPOLES FOR CORRECTION OF CHROMATIC
ABERRATIONS IN BEAM LINES WITH HORIZONTAL AND VERTICAL DISPERSIONS
\vspace{-0.5cm}}
\author{V. Balandin, W. Decking, N. Golubeva\thanks{nina.golubeva@desy.de} \\
DESY, Hamburg, Germany}

\maketitle

\begin{abstract}
In this article we discuss the usage of tilted multipoles 
for correction of chromatic aberrations in the design of 
the beam switchyard arc at the European X-Ray
Free-Electron Laser (XFEL) Facility~\cite{XFEL}.
\end{abstract}

\section{INTRODUCTION}

The European XFEL has been planed as a multiuser facility and from the
beginning will have the possibility to distribute electron bunches of one 
beam pulse to one or the other of two electron beamlines, each serving 
its own set of undulators. Additional space is reserved for 
the later addition of a third electron beamline.
Because different users have contradictory requirements
to the bunch repetition pattern, 
operational flexibility will be reached by a distribution system
which will use very stable flat-top kickers for directing beam into 
the undulator beamlines and fast single bunch kickers to kick
individual bunches into the transport line to the beam dump
before the beam distribution~\cite{XFEL,Winni}.

Both, the beam separation between undulator beamlines and beam deflection 
into the beam dump will be realized with a kicker-septum scheme.
While the beam quality in the dump line is not an issue, 
the optics of the beam separation between two undulator beamlines
must meet a very tight set of performance specifications.
It should be able to accept bunches with different energies 
(up to $\pm1.5\%$ from nominal energy)
and transport them without any noticeable deterioration
not only transverse, but also longitudinal beam parameters, i.e.
it must be sufficiently achromatic and sufficiently isochronous.
Besides that it is necessary to avoid magnet collisions in the design,
and to keep the degradation of the beam quality due to collective effects
within acceptable limits.

In this paper we discuss the optics solution for the beam separation area  
between two undulator beamlines (see Fig.1)
with the main attention played to the improvement of the
chromatic properties of the beam deflection arc by usage of sextupole and octupole 
magnets. Because of the Lambertson type septums used in the design,
the deflection arc has nonzero horizontal and vertical dispersions 
simultaneously. This means that regardless of the fact that 
the linear on-energy betatron motion is still transversely uncoupled in such a beamline,
we have not only the non-linear dispersions generated in both transverse planes,
but also vertical and horizontal oscillations become chromatically coupled 
due to vertical dispersion in the horizontal bending magnets
and horizontal dispersion in the vertical dipoles.
Nevertheless, because these effects are not the result of magnet 
misalignments and imperfections and are well controlled by the linear
optics design, the usage of tilted sextupoles and octupoles in such a beamline
allows to maintain the total number of multipoles 
required for correction of chromatic aberrations 
on the same level as required in the mid-plane symmetric systems.  

\noindent
\begin{figure}[ht]
    \centering
    \includegraphics*[width=80mm]{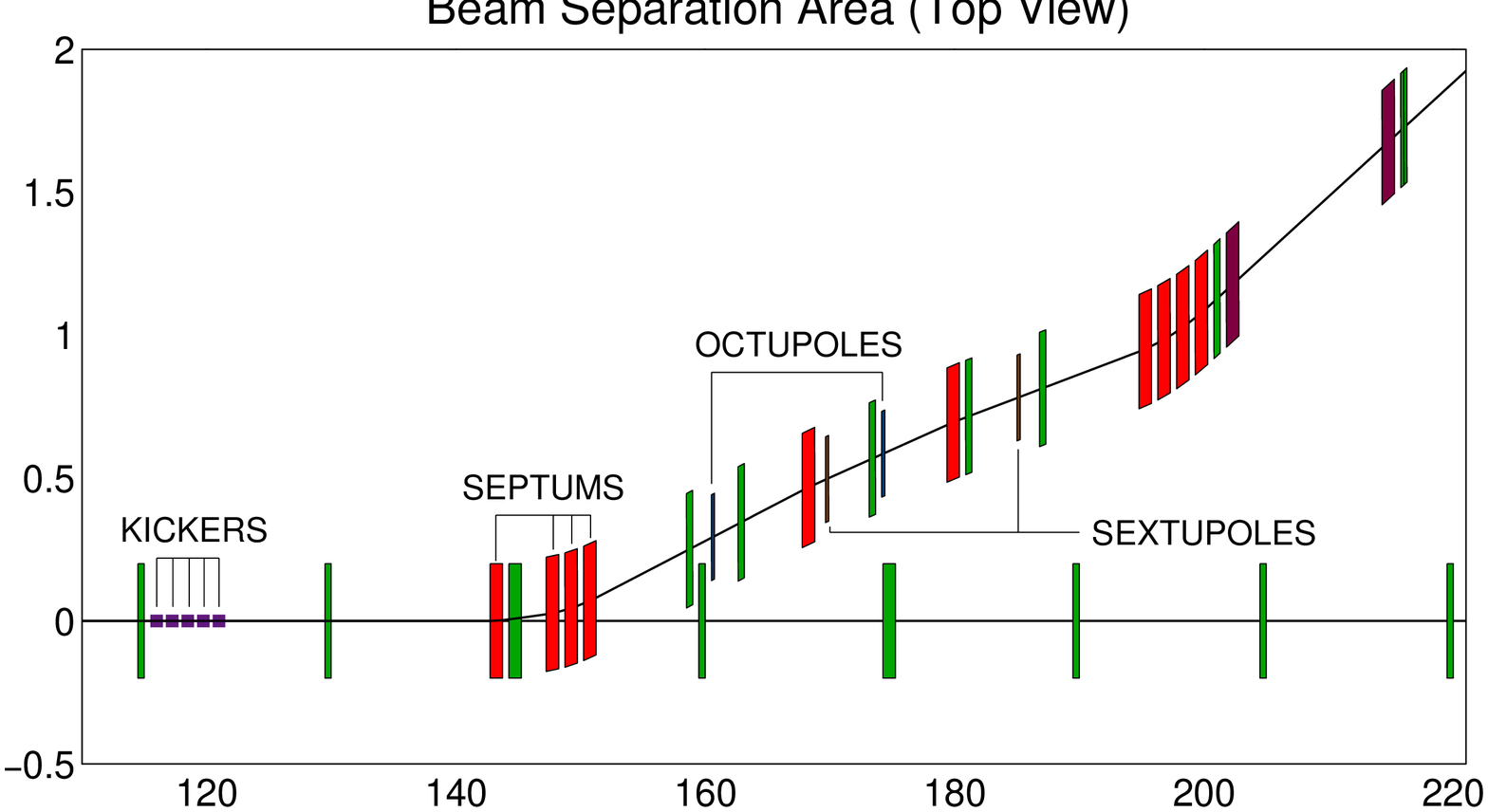}
    \vspace{-0.2cm}
    \caption{Top view of the separation area between two electron
    beamlines.
    Green, red and purple colors mark quadrupole
    magnets, and horizontal and vertical dipole magnets, respectively.
    Horizontal and vertical distances are measured in meters.}
    \label{fig1}
\end{figure}

\vspace{-0.5cm}
\section{SECOND-ORDER 
CHROMATIC ABERRATIONS DUE TO SEXTUPOLES}

In this section we give formulas for the sextupole
contributions to the second-order chromatic aberrations
from which one can see similarities and differences in the usage
of tilted sextupole magnets in the beamlines with non-flat
dispersion and in beamlines which bend the beam 
only horizontally (formulas (\ref{chAbb_16})-(\ref{chAbb_30}) 
with arbitrary angle $\theta$ and with angle $\theta$
multiple of $180^{\circ}$, respectively).
The effect of octupoles can be calculated and
analyzed in a similar fashion and due to space limitation is
not given here.

As usual, we take the path length along the reference orbit $\tau$ 
to be the independent variable and use a complete set 
of symplectic variables
$\mbox{\boldmath $z$} = (x, p_x, y, p_y, \sigma, \varepsilon)$
as particle coordinates~\cite{MaisRipken, BG}.
In these variables the Hamiltonian describing the motion of a particle
in the magnetostatic system of interest can be written as

\noindent
\begin{eqnarray}
H(\mbox{\boldmath $z$}) \,=\, 
\varepsilon - (1 + h x + \alpha y) \cdot
\Big[ \Big((1 + \varepsilon)^2 
\Big. \Big.
\nonumber
\end{eqnarray}
\vspace{-0.8cm}
\noindent
\begin{eqnarray}
\Big. \Big.
- (p_x - \bar{A}_x)^2 - (p_y - \bar{A}_y)^2 -
(\varepsilon / \gamma_0)^2
\Big)^{1 / 2} +  \bar{A}_z \Big],
\label{Ham1}
\end{eqnarray}

\noindent
where $\bar{A}_x$, $\bar{A}_y$ and $\bar{A}_z$ are the components
of the magnetic vector potential multiplied by the rigidity
of the reference particle, 
and $h$ and $\alpha$ are the horizontal and vertical
curvatures of the reference orbit, respectively. We assume that 
$h(\tau)\cdot\alpha(\tau)\equiv0$
and that both curvatures are positive
if the reference orbit bends in the direction opposite to that of 
the corresponding coordinate axis.
With the assumption that all magnets in our system are multipoles of separate function type
and with appropriately chosen vector potential, the Hamiltonian (\ref{Ham1})
expanded up to third order in the variables $\mbox{\boldmath $z$}$ then takes the form
$H =_3 H_2 + H_3$, where

\noindent
\begin{eqnarray}
H_2 \,=\, (1 \,/\, 2)\, 
\big(p_x^2 + p_y^2 + \varepsilon^2 \,/\, \gamma_0^2\big)
\,-\, ( h x + \alpha y) \, \varepsilon
\nonumber
\end{eqnarray}

\vspace{-0.2cm}

\noindent
\begin{eqnarray} 
+\, (1\,/\,2) \,\big(h^2 + k_1\big)      \, x^2 
\,+\, (1\,/\,2) \,\big(\alpha^2 - k_1\big) \, y^2, 
\label{Ham3}
\end{eqnarray}

\vspace{-0.2cm}

\noindent
\begin{eqnarray}
H_3 =
(1\,/\,2)\,\big(h x + \alpha y - \varepsilon\big)\, 
\big(p_x^2 + p_y^2 + \varepsilon^2\,/\,\gamma_0^2\big)
\nonumber
\end{eqnarray}

\vspace{-0.2cm}

\noindent
\begin{eqnarray}
-\,
(1\,/\,2)\,\big(h^{\prime} p_x - \alpha^{\prime} p_y\big)\, \big(x^2 - y^2\big) 
\nonumber
\end{eqnarray}

\vspace{-0.2cm}

\noindent
\begin{eqnarray}
-\,(1\,/\,6) \big(h^{\prime \prime}\, x^3
+ \alpha^{\prime \prime} \, y^3 \big)
\,+\,
k_2 \, V_s,
\label{Ham4}
\end{eqnarray}

\vspace{-0.2cm}

\noindent
\begin{eqnarray}
V_s \,=\,
\cos(3\varphi) \,\frac{x^3 - 3 xy^2}{6}
\,-\,\sin(3\varphi)\, \frac{y^3 - 3 x^2y}{6},
\label{Vs}
\end{eqnarray}

\noindent
and $=_n$ means equality up to order $n$, prime denotes 
differentiation with respect to the variable $\tau$, 
$\varphi$ is a sextupole tilt angle, and
$k_1$ and $k_2$ are quadrupole and sextupole coefficients,
respectively. 

We represent particle passage through our system
by a symplectic map $\,{\cal M}\,$ that maps the dynamical variables
$\,\mbox{\boldmath $z$}\,$ from the location $\tau = 0$
to the location $\tau = l$ and use for this map the following
Lie factorization

\noindent
\begin{eqnarray}
:{\cal M}: \, =_2 \, \exp(:{\cal F}_3(\mbox{\boldmath $z$}):) \, :M(l):.
\label{ForwardMap}
\end{eqnarray}

\noindent
Here $M(\tau) = (r_{km}(\tau))$ is a fundamental matrix solution of the linearized
system driven by the Hamiltonian (\ref{Ham3}) 
and the function ${\cal F}_3$ is a third order homogeneous polynomial:

\noindent
\begin{eqnarray}
{\cal F}_3(\mbox{\boldmath $z$}) \,=\, - \int_0^{l}  
H_3(\tau, \, M(\tau) \cdot \mbox{\boldmath $z$}) \, d \tau .
\label{Abb_1}
\end{eqnarray}

\noindent
We separate the polynomial ${\cal F}_3$
in two parts ${\cal F}_3 = {\cal F}_3^o + {\cal F}_3^s$,
where ${\cal F}_3^s$ describes the sextupole effects, and
use a notation $c_{abcde} ({\cal F}_3^s)$ for
the coefficient with which the monomial
$\,x^a \,p_x^b \,y^c \,p_y^d\, \varepsilon^e\,$ 
enters the polynomial ${\cal F}_3^s$.
Using polar coordinates $r_D$ and $\theta$
for the $r_{16}$ and $r_{36}$ elements, namely taking
$r_{16} = r_D \cos(\theta)$ and $r_{36} = r_D \sin(\theta)$,
the formulas 
for the sextupole contributions to the chromatic aberrations 
can be written as follows.

\paragraph{Chromatic Coupling Terms}

\noindent
\begin{eqnarray}
c_{10101}({\cal F}_3^s) \,=\,
\int_0^{l}
k_2 \,r_{11} \,r _{33}
\,r_D\,\sin(\theta - 3 \varphi) \, d \tau,
\label{chAbb_16}
\end{eqnarray}

\vspace{-0.4cm}

\noindent
\begin{eqnarray}
c_{01011}({\cal F}_3^s) \,=\,
\int_0^{l}
k_2 \,r_{12} \,r _{34}
\,r_D\,\sin(\theta - 3 \varphi) \, d \tau,
\label{chAbb_17}
\end{eqnarray}

\vspace{-0.4cm}

\noindent
\begin{eqnarray}
c_{10011}({\cal F}_3^s) \,=\,
\int_0^{l}
k_2 \,r_{11} \,r _{34}
\,r_D\,\sin(\theta - 3 \varphi) \, d \tau,
\label{chAbb_18}
\end{eqnarray}

\vspace{-0.4cm}

\noindent
\begin{eqnarray}
c_{01101}({\cal F}_3^s) \,=\,
\int_0^{l}
k_2 \,r_{12} \,r _{33}
\,r_D\,\sin(\theta - 3 \varphi) \, d \tau.
\label{chAbb_19}
\end{eqnarray}

\paragraph{Chromatic Focusing Terms}

\noindent
\begin{eqnarray}
c_{20001}({\cal F}_3^s) \,=\,
-\frac{1}{2} \int_0^{l}
k_2 \,r_{11}^2 \,
r_D\,\cos(\theta - 3 \varphi) \, d \tau,
\label{chAbb_20}
\end{eqnarray}

\vspace{-0.4cm}

\noindent
\begin{eqnarray}
c_{02001}({\cal F}_3^s) \,=\,
-\frac{1}{2} \int_0^{l}
k_2 \,r_{12}^2 \,
r_D\,\cos(\theta - 3 \varphi) \, d \tau,
\label{chAbb_21}
\end{eqnarray}

\vspace{-0.4cm}

\noindent
\begin{eqnarray}
c_{11001}({\cal F}_3^s) \,=\,
-\int_0^{l}
k_2 \,r_{11}\,r_{12} \,
r_D\,\cos(\theta - 3 \varphi) \, d \tau,
\label{chAbb_22}
\end{eqnarray}

\vspace{-0.4cm}

\noindent
\begin{eqnarray}
c_{00201}({\cal F}_3^s) \,=\,
\frac{1}{2} \int_0^{l}
k_2 \,r_{33}^2 \,
r_D\,\cos(\theta - 3 \varphi) \, d \tau,
\label{chAbb_23}
\end{eqnarray}

\vspace{-0.4cm}

\noindent
\begin{eqnarray}
c_{00021}({\cal F}_3^s) \,=\,
\frac{1}{2} \int_0^{l}
k_2 \,r_{34}^2 \,
r_D\,\cos(\theta - 3 \varphi) \, d \tau,
\label{chAbb_24}
\end{eqnarray}

\vspace{-0.4cm}

\noindent
\begin{eqnarray}
c_{00111}({\cal F}_3^s) \,=\,
\int_0^{l}
k_2 \,r_{33}\,r_{34}\,
r_D\,\cos(\theta - 3 \varphi) \, d \tau.
\label{chAbb_25}
\end{eqnarray}

\paragraph{Terms Responsible for the Second Order\\ 
Transverse and Longitudinal Dispersions}

\noindent
\begin{eqnarray}
c_{10002}({\cal F}_3^s) \,=\,
-\frac{1}{2} \int_0^{l}
k_2 \,r_{11} \,
r_D^2\,\cos(2 \theta - 3 \varphi) \, d \tau, 
\label{chAbb_26}
\end{eqnarray}

\vspace{-0.4cm}

\noindent
\begin{eqnarray}
c_{01002}({\cal F}_3^s) \,=\,
-\frac{1}{2}\int_0^{l}
k_2 \,r_{12}\,
r_D^2\,\cos(2 \theta - 3 \varphi) \, d \tau, 
\label{chAbb_27}
\end{eqnarray}

\vspace{-0.4cm}

\noindent
\begin{eqnarray}
c_{00102}({\cal F}_3^s) \,=\,
\frac{1}{2} \int_0^{l}
k_2 \,r_{33} \,
r_D^2\,\sin(2 \theta - 3 \varphi) \, d \tau, 
\label{chAbb_28}
\end{eqnarray}

\vspace{-0.4cm}

\noindent
\begin{eqnarray}
c_{00012}({\cal F}_3^s) \,=\,
\frac{1}{2}\int_0^{l}
k_2 \,r_{34}\,
r_D^2\,\sin(2 \theta - 3 \varphi) \, d \tau.
\label{chAbb_29}
\end{eqnarray}

\vspace{-0.4cm}

\noindent
\begin{eqnarray}
c_{00003}({\cal F}_3^s) \,=\,
-\frac{1}{6} \int_0^{l}
k_2 \,r_D^3\,
\cos(3 \theta - 3 \varphi) 
\, d \tau.
\label{chAbb_30}
\end{eqnarray}

\section{BEAM DEFLECTION ARC}

The beam deflection arc starts from
the kickers which deflect beam vertically
and, after enhancement of this deflection by the
following quadrupole, the beam arrives at the 
entrance of the first Lambertson septum magnet
with the vertical separation from the horizontal midplane
$y=0$ of about $18\,mm$. The first septum magnet is tilted by
approximately $11^{\circ}$ in such a way that it bends
particles not only horizontally but also slightly upward.
It is done in order to compensate the downward deflection
produced by the vertically focusing large aperture quadrupole 
that follows after the septum, and in order to have the beam traveling
in parallel to the horizontal midplane at the entrance of the
three remaining (non-tilted) septum magnets, as can be seen in
Fig.1 and Fig.2. 
The rest of the deflection arc is constructed  
from ordinary multipoles and the arc ends by a dogleg
consisting of two vertical dipoles, which is used
for bringing beam back to the horizontal plane $y =0$ 
and for closing the linear vertical dispersion.
The $r_{56}$ coefficient of the transfer matrix of the
total deflection arc (considered from the entrance of
the first kicker up to the exit of the last vertical dipole)
is equal to zero, i.e. the deflection arc is a first-order
isochronous beamline. This is achieved by usage of two reverse
bend dipoles placed close to the arc center.
The entrance Twiss parameters of the deflection arc are fixed
and are defined by the behavior of the betatron functions in the
straight beamline. The exit Twiss functions are such that 
they allow easy matching to the periodic downstream transport
channel (see Fig.3). Two tilted sextupoles and two tilted octupoles
are placed in the arc to provide 
the required chromatic properties of the beam transport 
(see Fig.1 and Fig.4). Note that the optimization of the number
of sextupoles and octupoles, and their positions, strengths 
and tilt angles was not a separate task after the finishing of

\begin{figure}[t]
    \centering
    \includegraphics*[width=75mm]{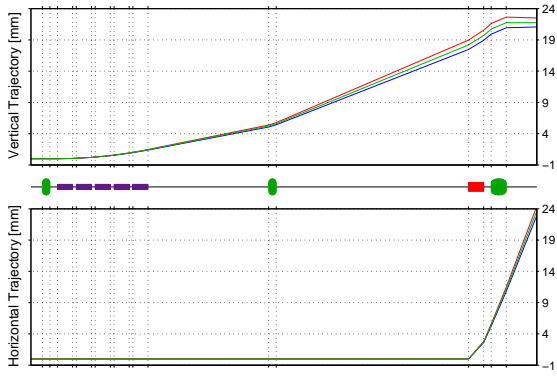}
    \vspace{-0.2cm}
    \caption{Trajectories of the kicked particles
    in the beginning of the separation area.
    The relative energy deviations are equal to
    $-3\%, \,0\%$ and $+3\%$ (red, green and blue curves,
    respectively).}
    \vspace{-0.5cm}
    \label{fig7}
\end{figure}

\begin{figure}[h]
    \centering
    \includegraphics*[width=75mm]{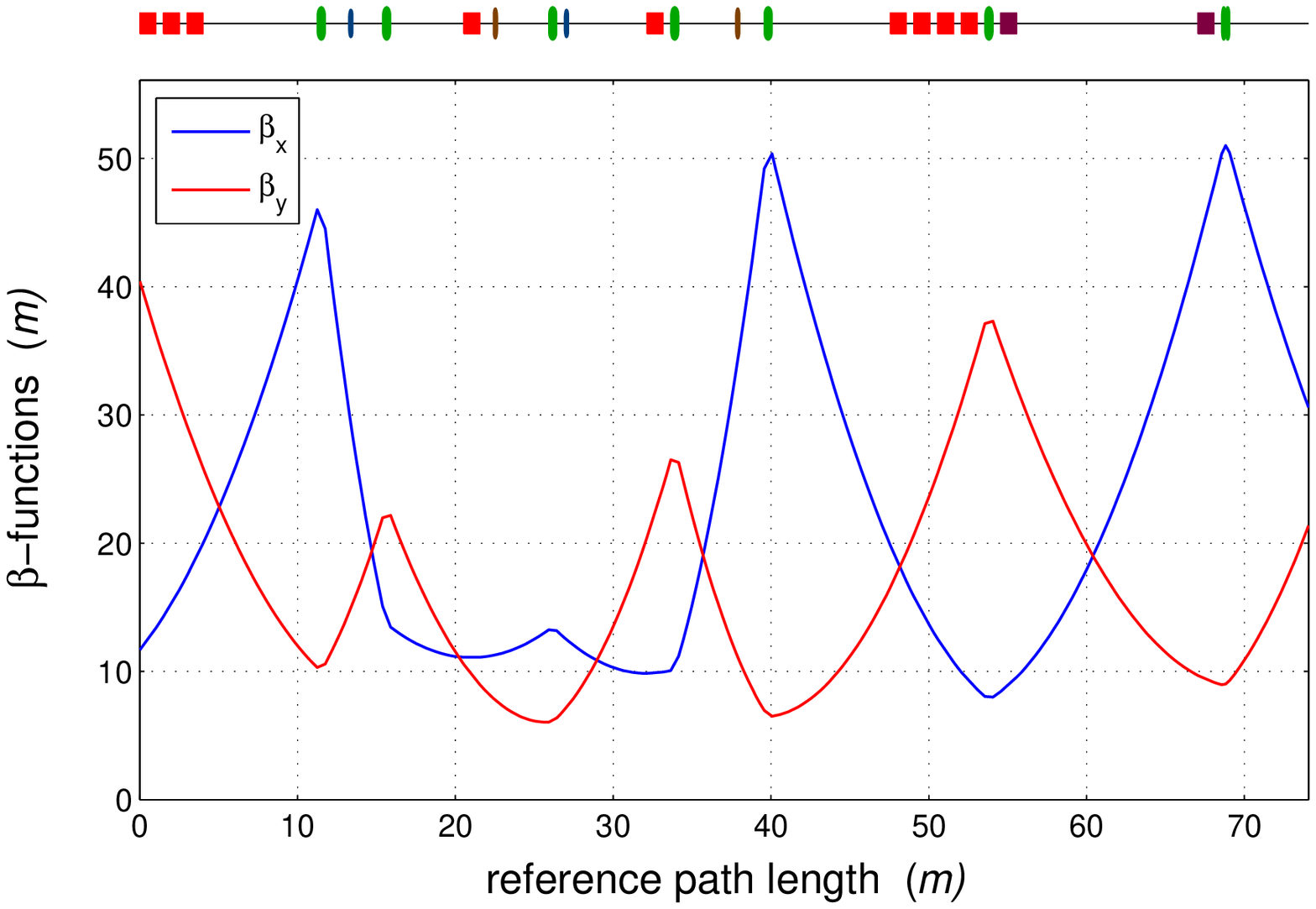}
    \includegraphics*[width=75mm]{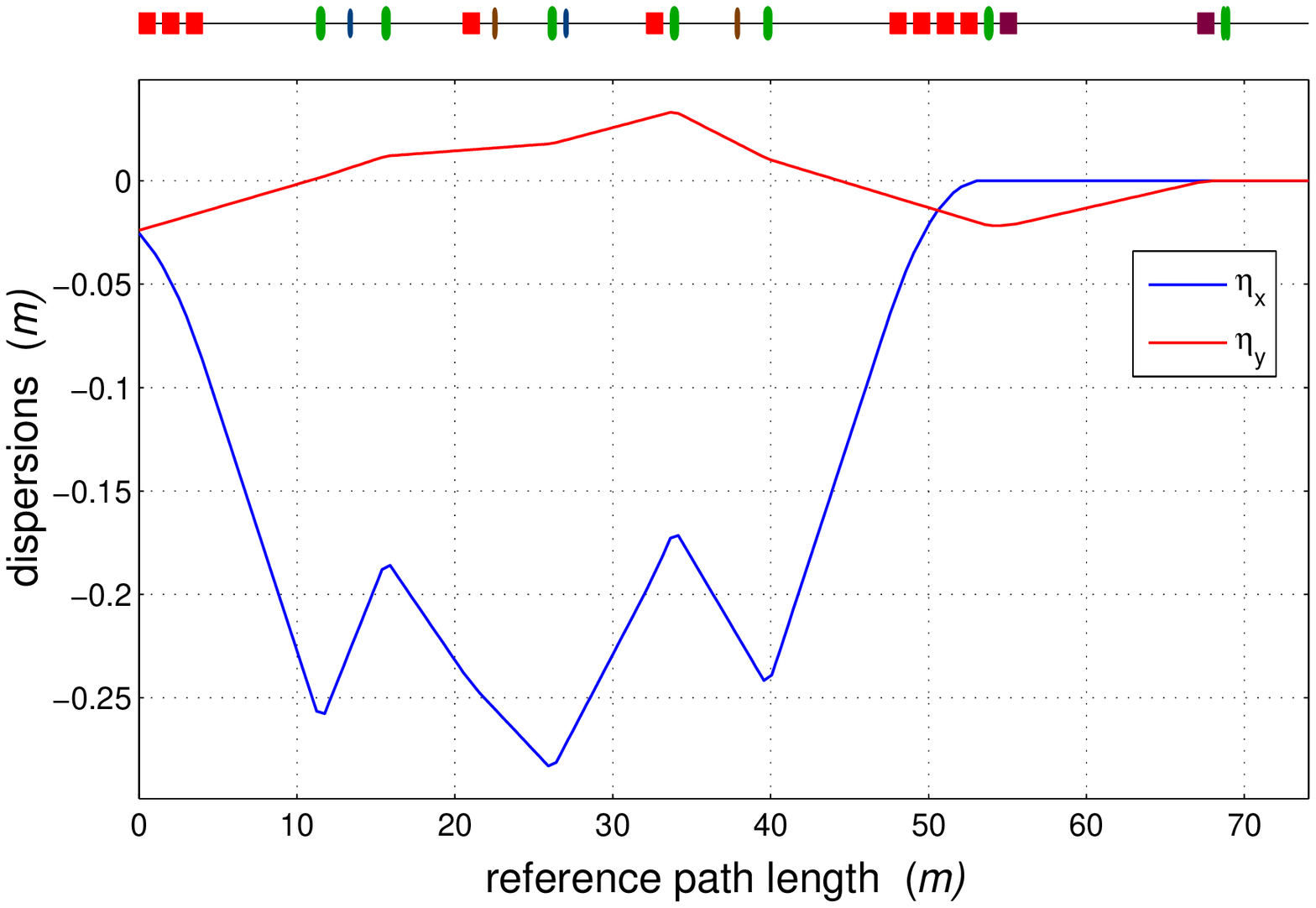}
    \vspace{-0.2cm}
    \caption{Betatron and dispersion functions along deflection arc
    shown starting from the entrance of the first non-tilted septum.}
    \label{fig2}
\end{figure}

\begin{figure}[t]
    \centering
    \includegraphics*[width=75mm]{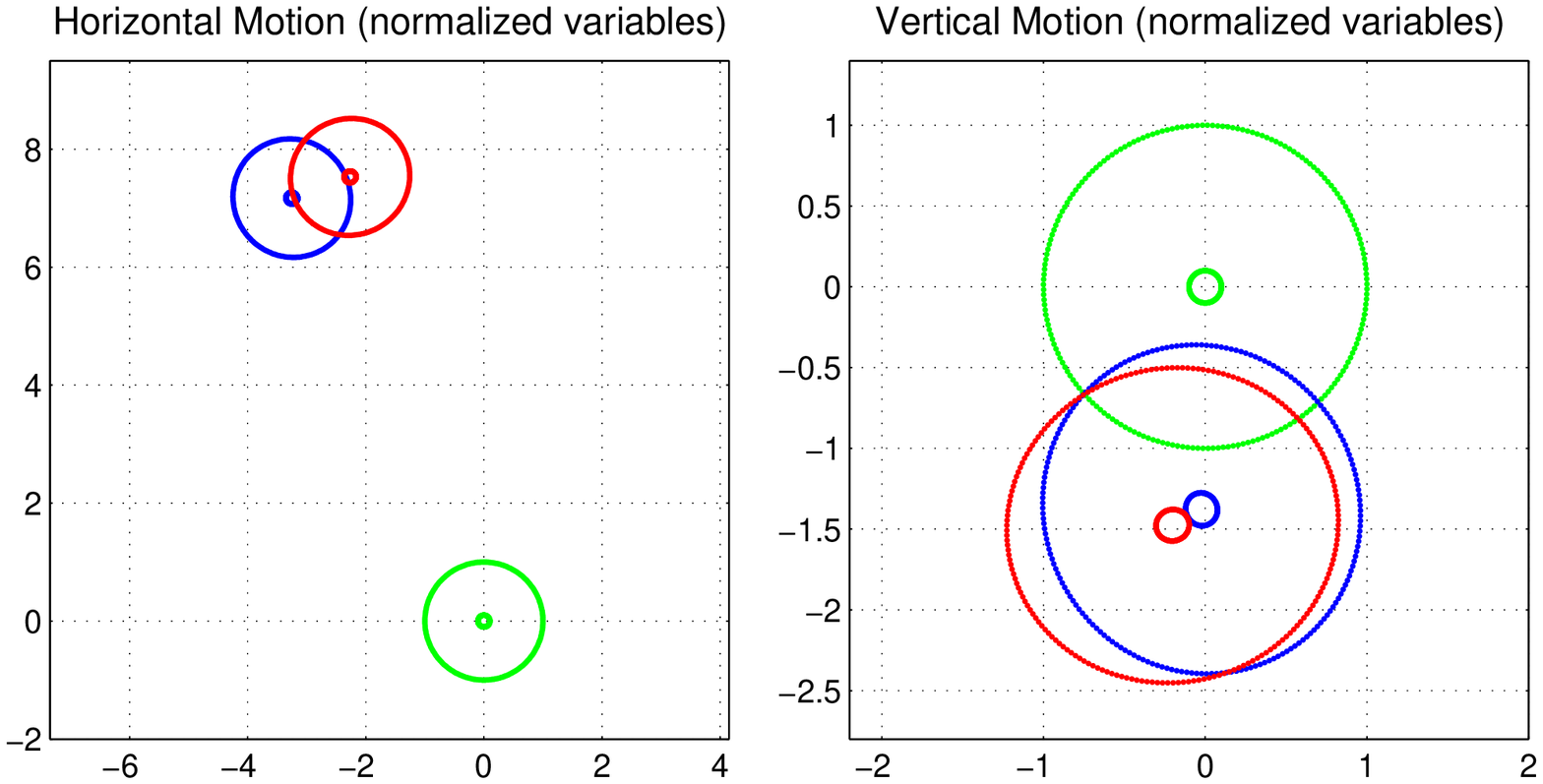}
    \includegraphics*[width=75mm]{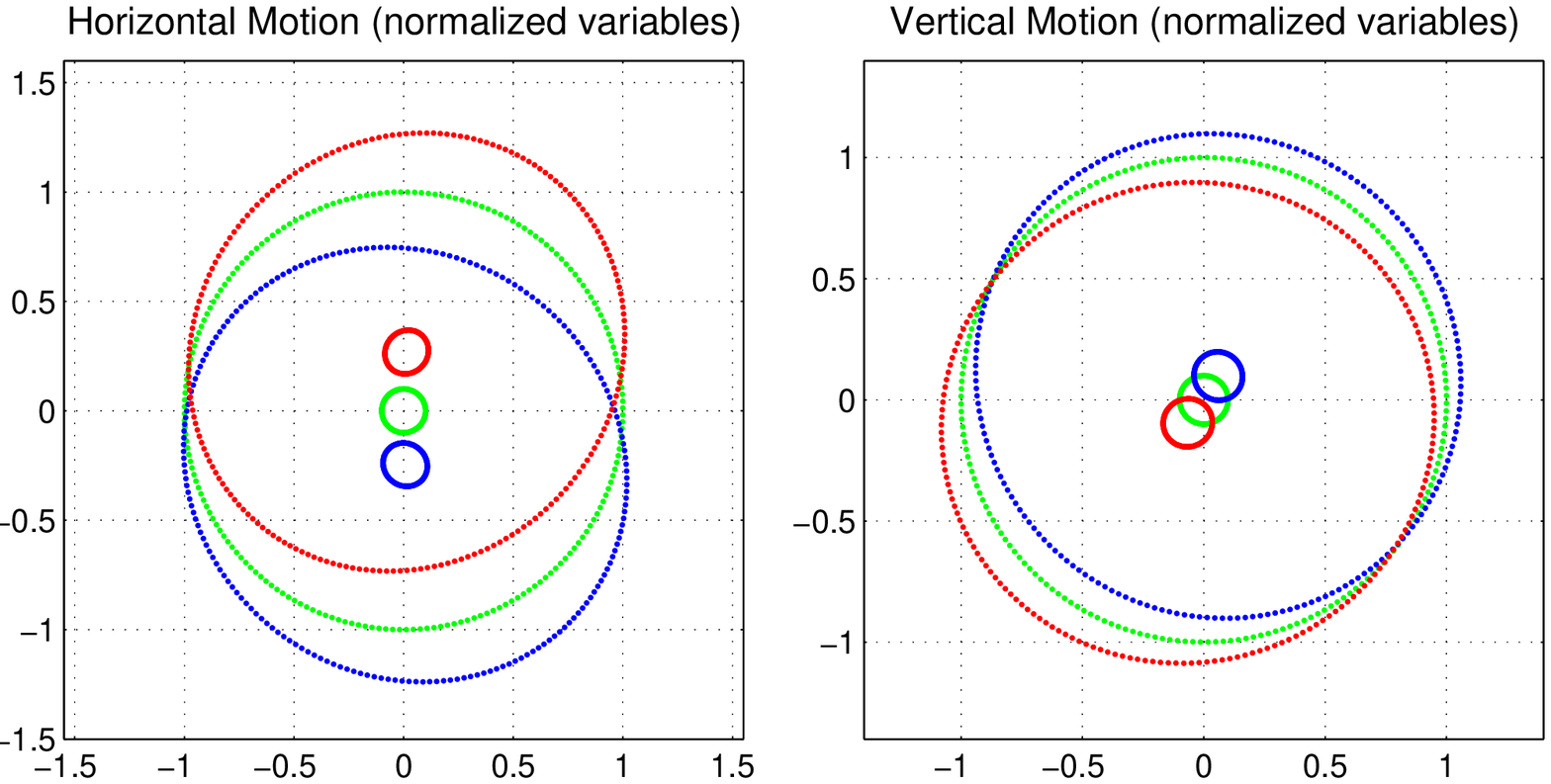}
    \includegraphics*[width=75mm]{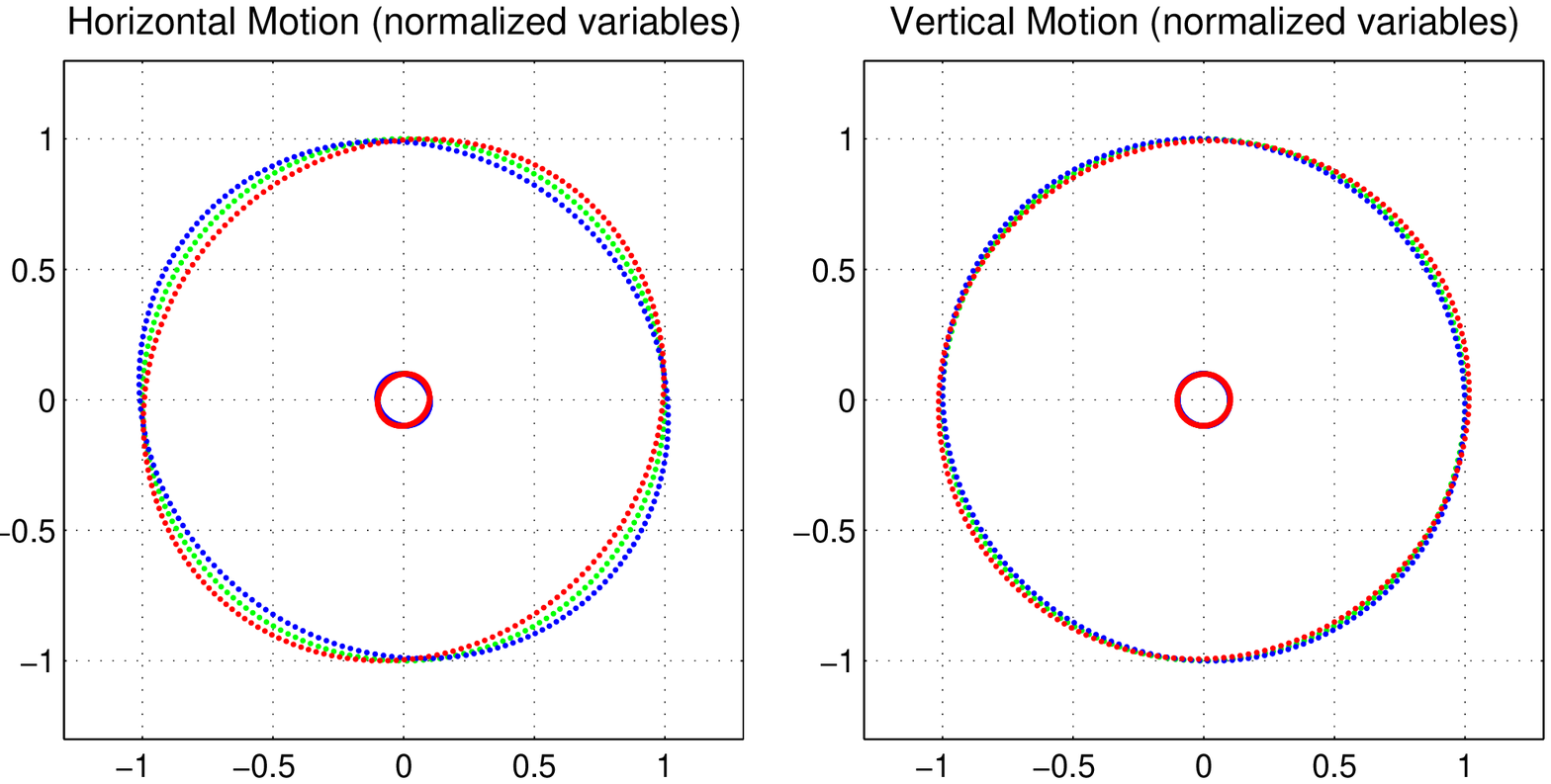}
    \vspace{-0.2cm}
    \caption{Phase space portraits of monochromatic 
    $0.1\sigma_{x,y}$ and $1\sigma_{x,y}$ ellipses
    (matched at the entrance) after tracking through
    the deflection arc. 
    The relative energy deviations are equal to
    $-1.5\%, \,0\%$ and $+1.5\%$ (red, green and blue ellipses,
    respectively).
    Sextupoles and octupoles are switched off (upper plots),
    sextupoles are on and octupoles are off (middle plots),
    sextupoles and octupoles are on (lower plots).}
    \vspace{-0.3cm}
    \label{fig4}
\end{figure}

\noindent
the design of the linear optics, but both, linear and non-linear optics
were designed together.

The arc design presented in this paper meets all design specifications
from the point of view of single particle beam dynamics.
The impact of collective effects on the beam quality still requires
additional investigations~\cite{Dohlus}.

\end{document}